\documentclass[a4paper]{article}

\usepackage{INTERSPEECH2022}

\title{Improving Speech Emotion Recognition Through \\ Focus and Calibration Attention Mechanisms}
\name{Junghun Kim, Yoojin An and Jihie Kim}
\address{Department of Artificial Intelligence,
Dongguk University, Seoul, Korea}
\email{\{riseup32, insuwon9\}@dongguk.edu, jihie.kim@dgu.edu}

\begin{document}

\maketitle
\begin{abstract}
Attention has become one of the most commonly used mechanisms in deep learning approaches. The attention mechanism can help the system focus more on the feature space’s critical regions. For example, high amplitude regions can play an important role for Speech Emotion Recognition (SER). In this paper, we identify misalignments between the attention and the signal amplitude in the existing multi-head self-attention. To improve the attention area, we propose to use a Focus-Attention (FA) mechanism and a novel Calibration-Attention (CA) mechanism in combination with the multi-head self-attention. Through the FA mechanism, the network can detect the largest amplitude part in the segment. By employing the CA mechanism, the network can modulate the information flow by assigning different weights to each attention head and improve the utilization of surrounding contexts. To evaluate the proposed method, experiments are performed with the IEMOCAP and RAVDESS datasets. Experimental results show that the proposed framework significantly outperforms the state-of-the-art approaches on both datasets.
\end{abstract}
\noindent\textbf{Index Terms}: speech recognition, emotion, attention

\section{Introduction}

The field of speech recognition has been spotlighted as a promising field of research with the rapid development of Automatic Speech Recognition (ASR). In particular, Speech Emotion Recognition (SER) is one of the most critical technologies for criminal investigation, medical treatment, real-time sentiment analysis, etc. In light-weight SER applications such as real-time emotion recognition on mobile devices, models with speech modality only are more desirable than heavier multi-modal (text and speech) models. We aim at improving SER accuracy for such tasks. Feature sets appropriated for SER have been researched and used in INTERSPEECH 2009 (IS09) Emotion Challenge \cite{C1}, AVEC Challenge \cite{C2} and extended Geneva Minimalistic Acoustic Parameter Set (eGeMAPS) \cite{C3}. These feature sets are usually hand-crafted Low-Level Descriptors (LLDs). However, thanks to the development of deep learning models, recent studies \cite{C4, C5, C6, C7} effectively use task-specific features extracted directly from the amplitude-related features such as spectrograms or Mel Frequency Cepstral Coefficients (MFCCs). Following the recent studies, we use MFCCs for our end-to-end approach.

Although deep learning approaches provide good performance, the results are difficult to understand or explain as they provide black-box models. A way to mitigate the shortcoming is the use of attention mechanisms. The attention maps extracted from the reasoning process help understand the attention distribution across the feature space. Also, the attention mechanism can improve the performance by focusing more on the feature space's critical regions. Since images and texts can be intuitively analyzed with attention distribution, many studies in computer vision and natural language processing \cite{C8, C9, C10, C11} have been conducted to improve attention performance by focusing more on the critical regions and adjusting misalignments. However, the critical regions in speech signals are not as intuitive as images or texts, making it difficult to understand and analyze the attention maps. In SER, \cite{C12} showed that the highest amplitude plays an important role in emotion recognition performance because each emotion has a different decibel level. Based on the study, we hypothesized that detecting the largest amplitude and learning the associated representations can improve emotion recognition performance. In experiments, we evaluate the hypothesis by visualizing the attention map and showing the performance improvement based on the attention alignment adjustment.

We first explore whether multi-head self-attention \cite{C13}, one of the most commonly used attention mechanisms in speech recognition, detects the critical region (the highest amplitude) in SER. In the exploration, we identify misalignments between the attention and the amplitude through a visualization of the attention map in the existing multi-head self-attention. To alleviate the attention maps' misalignment and improve the performance, in this paper, we propose a Focus-Attention (FA) mechanism and a novel Calibration-Attention (CA) mechanism. The FA mechanism used in the document summary task \cite{C14} employs a Gaussian distribution to focus on the areas with salient information. In this work, FA plays a role in detecting the largest amplitude part in the segment. There are multiple attention alignments due to the multi-head self-attention, and not all attention heads have an appropriate alignment. We propose the new CA mechanism to modulate the information flow by assigning different weights to each attention head and improve the utilization of surrounding contexts.

In the experiments, we use IEMOCAP \cite{C15} and RAVDESS \cite{C16} datasets, which are popular SER benchmark datasets, and we demonstrate that our approach achieves state-of-the-art results. Then we visualize the proposed methods' attention map and demonstrate that adjusting misalignments can improve SER performance. 

Our contributions are as follows:
\begin{itemize}
	\item We apply a focus-attention mechanism in SER to detect the largest amplitude part in the segment.
	\item We propose a novel calibration-attention mechanism to modulate the information flow and improve the utilization of surrounding contexts.
	\item In experiments, we achieve state-of-the-art performance with significant improvement over the existing state-of-the-art approaches on both IEMOCAP and RAVDESS datasets.
\end{itemize}

\section{Related Work}

\subsection{Features for SER}
The recent development of deep learning architectures has led us to use spectrograms or audio features such as MFCCs rather than hand-crafted LLDs. \cite{C6} extracted features directly from speech spectrogram instead of using traditional hand-crafted features to represent the emotion better. Such extracted features are used as the input of Convolutional Neural Networks (CNNs). \cite{C7} used both Mel-spectrograms and MFCCs for a dual-level model that contains two independent neural networks. They solved the time-frequency trade-off by extracting two different Mel-spectrograms having different window sizes. Also, they have shown that Mel-spectrograms are more challenging to learn than MFCCs through experiments. Our current work makes use of MFCCs.

\subsection{SER Models} 
At present, an attention mechanism is being used in almost every deep learning application. The attention mechanism can improve the performance by focusing more on the feature space's critical regions. Attentions can also mitigate the black box problem, which is a disadvantage of the deep learning model structure, to some extent. In the field of speech recognition, the attention mechanism has been used in tasks such as ASR \cite{C17, C18} and SER \cite{C19, C20, C21}. In particular, after \cite{C13} showed a strong performance of self-attention in the machine translation task, self-attention was also attempted to assign more weight to the critical regions in the SER task \cite{C6, C22} and proved its effectiveness. \cite{C6} used the self-attention mechanism to focus on the emotion that appears in a particular part, not in the whole utterance. \cite{C22} combined Dilated Residual Network (DRN) and multi-head self-attention to enhance the importing of emotion-salient information, and we denote this model as DRN-MHSA in this paper. Other state-of-the-art models in SER are as follows. Audio-BRE \cite{C23} uses a Bidirectional Recurrent Encoder (BRE) model based on the Long Short-Term Memory (LSTM) using MFCCs. A-DCNN \cite{C24} uses a SincNet filter layer \cite{C25} to learn custom filter banks from speech audio. Unlike \cite{C6, C22} using existing self-attention, we propose an improved self-attention using FA and CA mechanisms and verify the assumption that detecting the largest amplitude and learning expressive representations improves the emotion recognition performance.

\section{Approach}

In this paper, we propose to use a Focus-Attention (FA) mechanism and a novel Calibration-Attention (CA) mechanism in combination with multi-head self-attention. The FA plays a role in detecting the salient information, and the CA can modulate the information. This model architecture can improve the self-attention by learning better attention distribution in SER. The overall architecture of our model is illustrated in Fig. 1.

   \begin{figure*}[th]
      \centering
      \includegraphics[width=0.65 \linewidth]{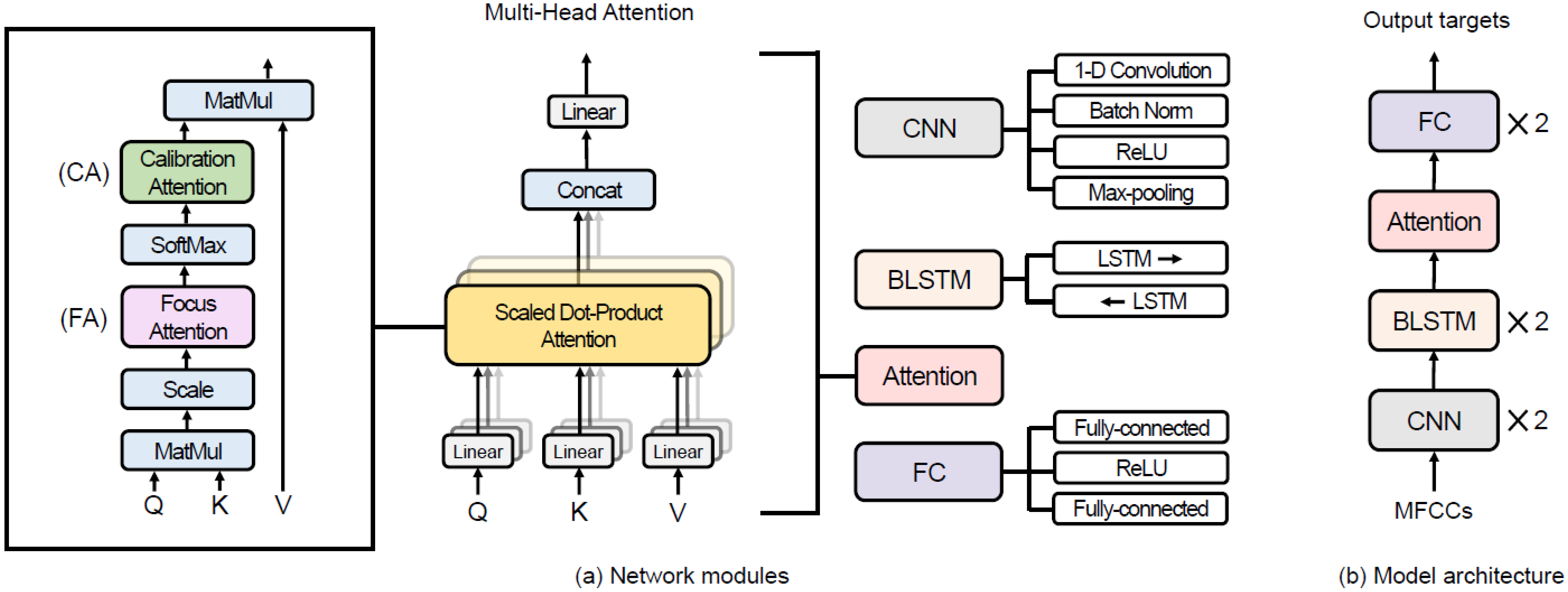}
      \caption{\begin{small}Network modules and the overall architecture.\end{small}}
   \end{figure*}

\subsection{Multi-Head Self-Attention}
As with the concept of \cite{C13}, the elements of the input sequences $X = (x_1, x_2, ..., x_m)$ are projected onto three different representations ($query$, $key$, and $value$) through a linear function. We define three representations as query $Q \in \mathbb{R}^{m \times d_m}$, key $K \in \mathbb{R}^{m \times d_m}$, and value $V \in \mathbb{R}^{m \times d_m}$. Self-attention is obtained by a scaled-dot product with query, key, and value:
$$
\ \mathrm{Attention}(Q, K, V) = \mathrm{SoftMax}(\frac{QK^T}{\sqrt{d_{m}}})V, \eqno{(1)}
$$
where $d_{m}$ is the dimension of linear projection output.

Instead of performing a self-attention once with $d_{m}$-dimensional query, key, and value, performing a self-attention $h$ times in parallel with $d_h$ = $d_{m}/h$ dimensional queries, keys, and values can jointly attend to information from different representation subspaces. Each calculated attention heads are concatenated and once again projected.
$$
\ \mathrm{MultiHead}(Q, K, V) = \mathrm{Concat}(H_1, ..., H_h)W^O,
$$
$$
\mathrm{where}\, H_i = \mathrm{Attention}(Q W_i^Q, K W_i^K, V W_i^V), \eqno{(2)}
$$
where $H_i \in \mathbb{R}^{d_m \times d_h}$ is $i$-{th} attention head, $W_i^Q \in \mathbb{R}^{d_m \times d_h}$, $W_i^K \in \mathbb{R}^{d_m \times d_h}$, $W_i^V \in \mathbb{R}^{d_m \times d_h}$ and $W^O \in \mathbb{R}^{h*d_h \times d_m}$ are the learnable weight matrices.

\subsection{Focus-Attention Mechanism}
A Focus-Attention (FA) mechanism was used within self-attention sub-layers to obtain salient information during encoding for the document summary task \cite{C14}. In our work, the FA mechanism detects the largest amplitude part as salient information in the segment during the encoding. This mechanism models a focal bias, which is the regularization term on attention score determined by the center position scalar and the coverage scope scalar. At the $i$-th sequence step, the center position scalar $\mu_i \in \mathbb{R}$ and the coverage scope scalar $\sigma_i \in \mathbb{R}$ are calculated through two linear projection processes:
$$
\ \mu _i=U_c^Ttanh(W_pQ_i+W_gG), \eqno{(3)}
$$
$$
\ \sigma  _i=U_d^Ttanh(W_pQ_i+W_gG), \eqno{(4)}
$$
where $W_p \in \mathbb{R}^{d_m \times d_m}$ and $W_g \in \mathbb{R}^{d_m \times d_m}$ are learnable shared weight matrices, $U_c \in \mathbb{R}^{d_m}$ and $U_d \in \mathbb{R}^{d_m}$ are learnable weight vectors. $G=\frac{1}{m} \sum_{i=1}^{m}Q_i \in \mathbb{R}^{d_m}$ is the mean vector providing complementary information. In addition, we regulate the $\mu _i$ and $\sigma  _i$ to values [0, m].
$$
\ \tilde{\mu _i}=m*\mathrm{Sigmoid}(\mu _i), \eqno{(5)}
$$
$$
\ \tilde{\sigma _i}=m*\mathrm{Sigmoid}(\sigma _i). \eqno{(6)}
$$

The focal bias $f_{i,j} \in \mathbb{R}$ is obtained and added to the attention before the softmax calculation.
$$
\ f_{i,j}=-\frac{(P_{j}-\tilde{\mu _i})^2}{(\tilde{\sigma _i})^2/2}, \eqno{(7)}
$$
$$
\ \mathrm{Attention}(Q, K, V) = \mathrm{SoftMax}(\frac{QK^T}{\sqrt{d_{m}}}\oplus f)V, \eqno{(8)}
$$
where $i, j \in \{1, 2, ..., m\}$. $P_{j} \in \mathbb{R}$ is the absolute position of the coefficient vector $x_j$ in the MFCCs and $\oplus$ denotes element-wise summation.

Moreover, we further adapt the FA mechanism into the multi-head manner as in Eq. 2.

\subsection{Calibration-Attention Mechanism}
There are multiple attention alignments due to the multi-head self-attention, and not all attention heads have an appropriate alignment. Therefore, we need to give more weight to the attention head with the appropriate alignment. We propose a novel Calibration-Attention (CA) mechanism to modulate the information flow by assigning different weights to each attention head and improve the utilization of surrounding contexts.

The following is the calibration process to modulate the information flow by obtaining calibration information $\mathbf{g} \in \mathbb{R}^{h}$ and calibration score $\mathbf{s} \in \mathbb{R}^{h}$. First, calibration information $\mathbf{g}$ is obtained by using Global Max Pooling (GMP) from the original attention head. When this calibration information $\mathbf{g}$ passes through Fully Connected (FC) layer using a nonlinear function, the calibration score $\mathbf{s}$ is obtained. This calibration score $\mathbf{s}$ has a value between 0 and 1. Finally, the information flow is modulated by multiplying the calibration score $\mathbf{s}$ with the original attention head.

$$
\ H^o = \mathrm{SoftMax}(\frac{QK^T}{\sqrt{d_{m}}}\oplus f), \eqno{(9)}
$$
$$
\ \mathbf{g} = \mathrm{GMP}(H^o), \eqno{(10)}
$$
$$
\ \mathbf{s} = \mathrm{Sigmoid}(W_{s}\mathbf{g}+b_{s}), \eqno{(11)}
$$
$$
\ H^s = \mathbf{s} \times H^o, \eqno{(12)}
$$
$$
\ \mathrm{Attention}(Q, K, V) = H^sV, \eqno{(13)}
$$
where $H^o \in \mathbb{R}^{h \times m \times m}$ and $H^s \in \mathbb{R}^{h \times m \times m}$ are attention head before / after calibration, respectively, $W_s \in \mathbb{R}^{h \times h}$ and $b_s \in \mathbb{R}^{h}$ are the weight matrix and bias vector.

   \begin{figure*}[th]
      \centering
      \includegraphics[width=1 \linewidth]{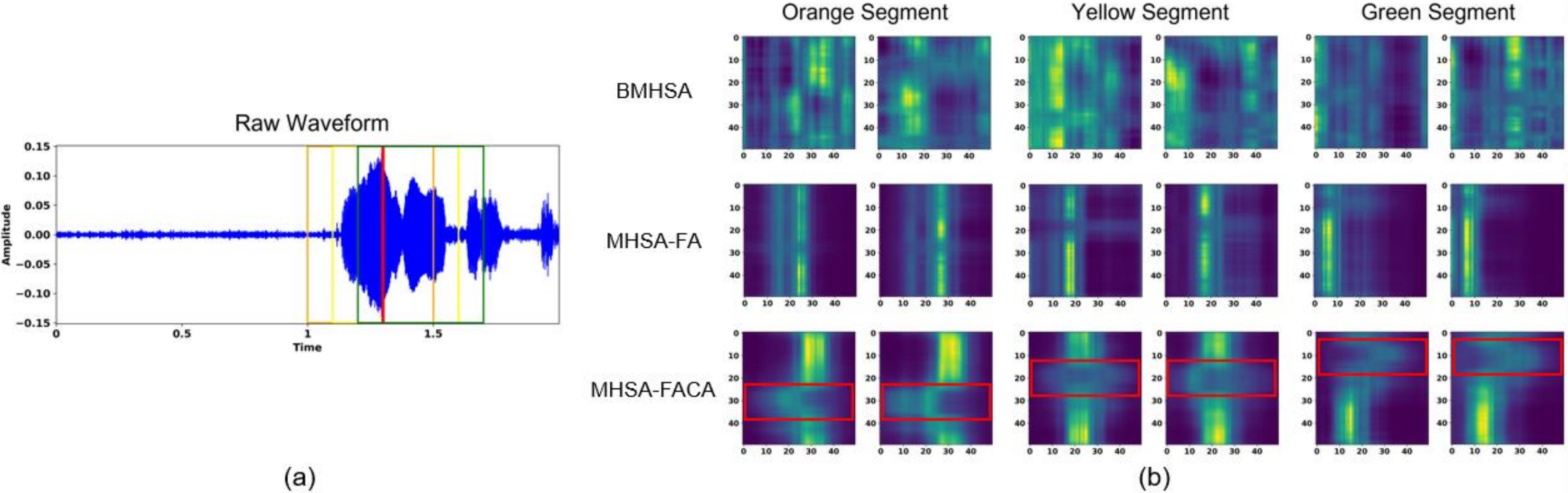}
      \caption{\begin{small}(a) is the raw waveform of the utterance. The utterance is divided into 0.5-second segments, with each segment marked in different colors (orange, yellow, or green). In the waveform, the part with the largest amplitude is marked with red line. (b) is the two heads' attention map for each segment, where light-colored regions indicate the attention that is prominent for the segment.\end{small}}
   \end{figure*}

\subsection{Model Architecture}
\cite{C26} proposed Convolutional, Long Short-Term Memory, Deep Neural Networks (CLDNNs), a combination of CNNs, LSTMs, and Deep Neural Networks (DNNs), and achieved better performance than any of those architectures individually. In CLDNNs, CNNs reduce the frequency variations, LSTMs model sequence features, and DNNs map features to distinguishable spaces. In this work, we use this model as the base structure and combine FA and CA as described above.

As shown in Fig. 1, first, we stack two 1D Convolutional layers with a kernel size of (3, 1) and a channel size of 64. Each Convolutional layer is followed by batch normalization and max-pooling with kernel and stride size (2, 1). Then, we stack two bidirectional LSTM layers with a hidden size of 256. The LSTM layer's output is connected by multi-head self-attention with a head size of 8, and finally, two FC layers are stacked. The FA mechanism is added before the softmax calculation in the scaled dot-product process, and CA mechanism operates before the multi-head is concatenated.

\section{Experiments}

\subsection{Dataset and Experimental Setup}

\textbf{Dataset.} We choose IEMOCAP \cite{C15} and RAVDESS \cite{C16} datasets, which are popular in SER, for the research of emotion classification. In IEMOCAP, following the previous research method, we add `exciting' class to happy class and use four classes \{happiness, anger, sadness, and neutral\}, including each \{1636, 1103, 1084, 1708\} utterances. RAVDESS contains 8 classes \{calmness, happiness, sadness, anger, fear, surprise, disgust, and neutral\}. To compare the performance with the previous approaches under the same condition, we use the same 10-fold and 5-fold cross-validation in IEMOCAP and RAVDESS, where 8, 1, 1 folds and 3, 1, 1 folds are train set, validation set, and test set, respectively.
\\
\textbf{Training.} Cross-entropy is employed as a loss function, and the Adam optimizer \cite{C27} with a learning rate of 3e-4 is employed. We train for 100 epochs and use a batch size of 128.
\\
\textbf{Metrics.} Weighted accuracy (WA) and unweighted accuracy (UA) are used to assess the model performance. Following the recent studies \cite{C22, C23, C25, C28, C29, C30, C31}, we use the averages from the 10-fold and 5-fold cross-validation as experimental results of IEMOCAP and RAVDESS, respectively.
\\
\textbf{Baselines.} The state-of-the-art models compared with our model are as follows: Audio-BRE \cite{C23}, DRN-MHSA \cite{C22}, A-DCNN \cite{C24}, and Audio-CNN-xvector \cite{C28} in IEMOCAP; Deep-CNN \cite{C29}, Head Fusion \cite{C30}, and QCNN \cite{C31} in RAVDESS.

\subsection{Performance evaluation}
We compare our model with the state-of-the-art models evaluated in the same metric setting (the most commonly used 10-fold and 5-fold cross-validation in IEMOCAP and RAVDESS, respectively) and speech modality. The basic multi-head self-attention model is denoted as BMHSA, the model with the FA mechanism added to BMHSA as MHSA-FA, and the model with CA mechanism added to MHSA-FA as MHSA-FACA. Table 1 shows the performance of MHSA-FACA compared with state-of-the-art models in IEMOCAP. Our model outperforms the state-of-the-art models by at least 2.21\% in WA and at least 4.43\% in UA. Table 2 shows comparison results on RAVDESS. MHSA-FACA outperforms the state-of-the-art models by at least 5.05\% in WA and at least 4.6\% in UA.

\subsection{Ablation Study}
We compare the performances according to the proposed module combination in IEMOCAP and show the results in Table 3. MHSA-FA improves performance by 1.9\% in WA and 1.78\% in UA over BMHSA. Also, MHSA-FACA brings additional performance improvements of 0.45\% in WA and 0.53\% in UA compared with MHSA-FA.

\begin{table}[t]
    \caption{\begin{small}Comparison results on IEMOCAP.\end{small}}
    \centering
{\small
  \begin{tabular}{ |c|c|c|c| } 
  \hline
  Model & WA(\%) & UA(\%) \\
  \hline
  audio-BRE \cite{C23} & 64.60 & 65.20 \\ 
  DRN-MHSA \cite{C22} & - & 67.40 \\ 
  A-DCNN \cite{C24} & 69.80 & - \\ 
  Audio-CNN-xvector \cite{C28} & 66.60 & 68.40 \\ 
  \hline
  MHSA-FACA (ours) & \textbf{72.01} & \textbf{72.83} \\ 
  \hline
  \end{tabular}
}
\end{table}

\begin{table}[t]
    \caption{\begin{small}Comparison results on RAVDESS.\end{small}}
    \centering
{\small
  \begin{tabular}{ |c|c|c|c| } 
  \hline
  Model & WA(\%) & UA(\%) \\
  \hline
  Deep-CNN \cite{C29} & - & 71.67 \\ 
  Head Fusion \cite{C30} & 77.80 & 77.40 \\ 
  QCNN \cite{C31} & - & 77.87 \\ 
  \hline
  MHSA-FACA (ours) & \textbf{82.85} & \textbf{82.47} \\ 
  \hline
  \end{tabular}
}
\end{table}

\begin{table}[t]
    \caption{\begin{small}Performances according to the proposed module combination in IEMOCAP.\end{small}}
    \centering
{\small
  \begin{tabular}{ |c|c|c|c| } 
  \hline
  Model & WA(\%) & UA(\%) \\
  \hline
  BMHSA (ours) & 69.66 & 70.52 \\
  MHSA-FA (ours) & 71.56 & 72.30 \\ 
  MHSA-FACA (ours) & \textbf{72.01} & \textbf{72.83} \\ 
  \hline
  \end{tabular}
}
\end{table}

To see how the proposed method changes attention distribution, the three models' attention maps are visualized and shown in Fig. 2. The utterances are divided into 0.5 sec (50 sequences in the attention map) segments. The segments have a hop of 0.1 sec (10 sequences in the attention map) for the sliding. The largest amplitude in the raw waveform in Fig. 2 (a) is represented by a red line, corresponding to in Fig. 2 (b) a sequence of 30 in the orange segment, 20 in the yellow segment, and 10 in the green segment, respectively. In the each attention map, the vertical direction means the query axis and the horizontal one is the key axis. BMHSA has failed to detect the largest amplitude and shows no particular pattern in the attention map. MHSA-FA detects the largest amplitude. Therefore, in each query, the attention is focused on the key corresponding to the largest amplitude. MHSA-FACA improves the utilization of surrounding contexts. If it is not a query of the largest amplitude, the attention is focused on the key corresponding to the largest amplitude. However, if it is a query of the largest amplitude, the attention is focused on the surrounding key instead of the largest amplitude, as shown as red boxes in Fig. 2 (b). Thus, MHSA-FACA seems to make use context information better in the inference process than MHSA-FA. As we predicted that detecting the largest amplitude and learning expressive representations improves emotion recognition performance, the model actually shows a significant performance improvement when it detects the largest amplitude. Also, when the model makes use of the surrounding context, it shows an additional performance improvement.

\section{Conclusions}

In this paper, we propose a focus-attention mechanism and a novel calibration-attention mechanism to improve self-attention in SER by alleviating the attention maps' misalignment. Using the focus-attention mechanism, the network can detect the largest amplitude part in the segment. Employing the calibration-attention mechanism, the network can modulate the information flow by assigning different weights to each attention head and improve the utilization of surrounding contexts. The proposed framework achieves state-of-the-art performance with significant improvement over the existing approaches.

\section{Acknowledgements}
This research was supported by the MSIT(Ministry of Science, ICT), Korea, under the High-Potential Individuals Global Training Program)(RS-2022-00155054) (50\%) and under the ITRC(Information Technology Research Center) support program(IITP-2022-2020-0-01789) (50\%), supervised by the IITP(Institute for Information \& Communications Technology Planning \& Evaluation).

\bibliographystyle{IEEEtran}

\bibliography{mybib}

\end{document}